\documentclass[twocolumn,english,prb]{revtex4-1}
\usepackage[latin9]{inputenc}
\usepackage{amsmath}
\usepackage{graphicx}
\usepackage{xcolor}

\makeatletter

\usepackage{babel}

\makeatother

\usepackage{babel}
\begin{document}

\title{Superconductor-like effects in an AC driven normal Mott-insulating
quantum dot array }

\author{Sanjeev Kumar}

\affiliation{Department of Theoretical Physics, Tata Institute of Fundamental
Research, Homi Bhabha Road, Navy Nagar, Mumbai 400005, India}

\author{Vikram Tripathi}

\affiliation{Department of Theoretical Physics, Tata Institute of Fundamental
Research, Homi Bhabha Road, Navy Nagar, Mumbai 400005, India}

\date{\today}
\begin{abstract}
We study the current response of an AC driven dissipative Mott insulator
system, a normal quantum dot array, using an analytical Keldysh field
theory approach. Deep in the Mott insulator regime, the nonequilibrium steady state (NESS) response resembles
a resistively shunted Josephson array, with a nonequilibrium Mott
insulating to conductor transition as the drive frequency $\Omega$
is increased. The diamagnetic component of the NESS in the conducting
phase is anomalous, implying negative inductance, strikingly reminiscent
of the $\eta$-pairing phase of a Josephson array with negative phase
stiffness. However in the presence of an additional DC field the signature
of supercurrent - Shapiro steps - is completely absent. We interpret
these properties as number-phase fluctuation effects shared with Josephson
systems rather than superconductivity.
\end{abstract}
\maketitle

\section{Introduction}
The nonequilibrium response of strongly correlated quantum systems
is a challenging problem requiring understanding of the many-body
excitation spectra, wavefunctions, dynamical bottlenecks and dissipative
processes. Driven Mott insulators are a prototypical example, exhibiting
diverse phenomena that are otherwise not present in their equilibrium
or linear-response regimes such as field/current driven insulator-metal
transitions \cite{PRB_met_insul_Werner_Golez,sakai_superconductivity_alkali,PRL_Eckstein_Werner,wall2011quantum,PhysRevB_Tsuji_Oka_Aoki},
Bloch oscillations or Wannier-Stark quantization\cite{S_Sankar_V_Tripathi,P_Werner_Floquet_dc_ac,Alexander_dc_Long_lived_noneqm_states,Kwon_dielectric_breakdown,Jop_Eckstein_Werner,oka_philipp_eckstein_Dielectric_Breakdown}
and current enhanced diamagnetism \cite{T_Oka_StrongDiamagnetism_Mott_insulator}.
A number of recent studies of optically excited Mott insulating half-filled
Hubbard models have proposed a new route to superconductivity through
doublon creation \cite{Olivier_ac_driven,philipp_eckstein_long_range_eta_pairing,Tomonori_Shirakawa_photoinduced_eta_pairing,werner_eckstein_entropy_doublonHolon,PRB_Yunoki,nature_pairing_due_to_ac,clark_enhancementOfpairing_ac,science_kota,T_Oka_P_Werner_ac_Repulsive_to_Attractive,PRL_Metastable_superfluidity_Rosch},
possibly an exotic $\eta$-pairing state \cite{yang_eta_pairing_hubbard_chain}.
The evidence comes from superconductor-like properties such as effective
attractive Coulomb correlations \cite{Olivier_ac_driven,clark_enhancementOfpairing_ac,T_Oka_P_Werner_ac_Repulsive_to_Attractive},
finite charge stiffness \cite{Millis2019optically} and off-diagonal
long range order parameter correlations \cite{Millis2019optically,PRL_Tindall_long_eta_pairing,Tomonori_Shirakawa_photoinduced_eta_pairing}.
However these half-filled Hubbard models are special since charge excitations necessarily create doublons. We therefore
ask whether superconductor-like properties may still
be seen in driven Mott insulators where charge excitations are not
associated with doublons. Specifically, we study the current response
of a dissipative Mott insulator system - an array of mesoscopic quantum
dots each with a large and arbitrary number of interacting electrons
- to an AC electric field quench.

We find that the NESS response deep in the Mott insulator regime has a striking resemblance with optically
driven resistively shunted Josephson arrays on either side of a superconductor-insulator
transition. We show that the frequency dependence of the current response
has regimes of both diamagnetic and insulating behavior as the AC
drive tunes the Mott insulator through a singularity separating insulator and metallic frequency dependences of the optical conductivity, indicating a nonequilibrium insulator-metal transition.
However the sign of the diamagnetic response is anomalous (negative),
tantamount to $\pi$-phase slips in the links, analogous to the $\eta$-pairing
phase of Josephson arrays with negative phase stiffness. In the presence
of a simultaneous DC bias, the DC $IV$ characteristics exhibit Josephson-like
photon-mediated tunneling in the form of current steps at bias values
separated by integer multiples of drive frequency $\hbar\Omega/e,$ but crucially, Shapiro steps - a key signature
of supercurrent - expected at integer multiples of $\hbar\Omega/2e$
are absent, unlike the observation in Josephson systems \cite{lankhorst2016giant_shapiro,matsuura2008CDW_shapiro,tinkham2004introduction}.
We propose that the similarities shared with reported optical response of Mott insulators are not on account of $\eta$-pairing or AC-induced superconductivity but are a manifestation of number-phase duality effects common to both. Strong charge fluctuations,
whether associated with underlying superconductivity or optical pumping,
suppress quantum fluctuations of the phases. 

Theoretical understanding of driven Mott insulators has received a
significant impetus by developments in the Keldysh dynamical
mean field theory (KDMFT) approach \cite{kdmft_hhg_Werner,peronaci2018resonant,tao_ness_ac,Alexander_dc_Long_lived_noneqm_states,Kwon_dielectric_breakdown,oka_philipp_eckstein_Dielectric_Breakdown,P_Schmidt_DMFT},
tensor network techniques \cite{PRL_Tindall_long_eta_pairing},
and analytic Bethe-ansatz \cite{oka_bethe_ansatz} including
the effective ${\cal PT}$-symmetric descriptions \cite{tripathi2016,fukui1998}.
Recently an alternate analytical large-$\mathcal{N}$ effective Keldysh
field theory approach \cite{S_Sankar_V_Tripathi} has been developed
based on the well-known Ambegaokar-Eckern-Schön (AES) rotor model
\cite{RevModPhys_aes2,PRL_AES} for electron transport in mesoscopic
quantum dot arrays, effectively a dissipative Mott insulator system.
This Keldysh formalism captures numerous nonequilibrium
DC phenomena including Bloch-like oscillations and the field-driven
insulator to metal transition.
It also provides an analytical treatment of the approach to the NESS.
Here we shall generalize this formalism for the AC response. 

The rest of the paper is organized as follows. In Sec. \ref{sec:model} we introduce the Keldysh AES model for transport in a quantum dot chain, and obtain an expression for the current response functional. Sec. \ref{sec:ACresponse} is devoted to the study of the current response to a uniform AC drive. The analysis not only confirms a number of results such as odd harmonic generation, Bloch-like oscillations and effective attractive local Coulomb correlations, hitherto obtained from numerical KDMFT studies, but also reveals some aspects missed in the numerical studies, most notably the slow decay of the Bloch-like oscillations. We find striking similarities to the current response of superconductor Josephson junction arrays with effectively negative superfluid stiffness, reminiscent of an $\eta$-pairing phase. To check if the superconductor-like optical response is indeed due to superconductivity in our system, we analyze in Sec. \ref{sec:shapiro} the current response when AC and DC driving fields are simultaneously present. Although photon mediated tunneling features, similar to superconducting Josephson arrays are also found here, the absence of Shapiro steps, a key signature of supercurrent, leads us to conclude that the properties are a manifestation of number-phase duality shared with superconductors. Sec. \ref{sec:discussion} contains a summary of our findings and a discussion.

\section{Model and formalism}\label{sec:model}

\begin{figure}
\includegraphics[width=1\columnwidth]{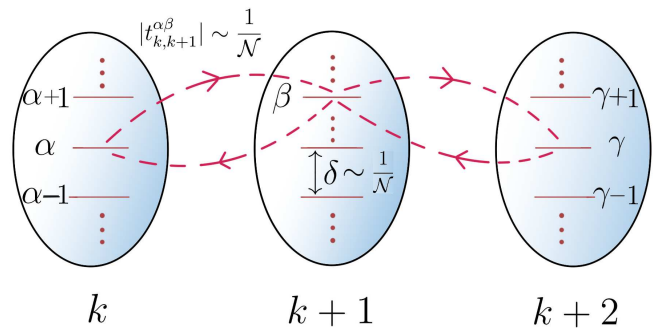}
\caption{\label{fig:largeN}Figure illustrating the tunneling processes between the normal quantum dots (labeled $k,$ $k+1,$ $k+2$) each of which contains a large number of single-electron levels $\epsilon_{k\alpha}.$ The mean level spacing $\delta \sim \mu/\mathcal{N},$ where $\mathcal{N} \gg 1$ is of the order of the number of conduction electrons in a dot, and $\mu$ is the bulk Fermi energy of the material of the dot. Transport is controlled by the dimensionless interdot tunneling conductance $g\sim |t|^2/\delta^2,$ where $|t|^2$ is the mean square of the distribution of the interdot tunneling elements, assumed to be a Gaussian white noise distribution $\langle t_{k,k+1}^{\alpha\beta}t_{k+1,k}^{\gamma\delta} \rangle = |t|^2\delta^{\alpha\delta}\delta^{\beta\gamma}$ for simplicity and concreteness. In order for $g$ to be physically meaningful, we require the tunneling elements $t_{k,k+1}^{\alpha\beta}$ to scale as $1/\mathcal{N}.$ The figure illustrates a higher order tunneling process involving three dots. The summation over the internal indices $\alpha,$ $\beta,$ $\gamma$ contributes a scaling factor $\mathcal{N}^3$ while the four tunneling elements contribute $1/\mathcal{N}^4.$ This higher order process thus scales as $1/\mathcal{N}$ and may be dropped in the large$\mathcal{N}$ limit.}
\end{figure}

Our starting point is a model Hamiltonian of a quantum dot array, 
\begin{equation}
H = H_0 + H_C + H_{\text{tun}}\label{eq:modelH},
\end{equation}
where 
\begin{equation}
H_0 = \sum_{k\alpha}(\epsilon_{k\alpha}-\mu)c^{\dagger}_{k\alpha}c_{k\alpha}
\end{equation}
 describes noninteracting electrons in the dots with energies $\epsilon_{k\alpha}$ for the $k^{\text{th}}$ dot,
 \begin{equation}
H_C = E_C\sum_{k}\left[\left(\sum_\alpha c^{\dagger}_{k\alpha}c_{k\alpha}\right)-N_{0}\right]^2
 \end{equation}
represents Coulomb interaction, and 
\begin{equation}
H_{\text{tun}}=\sum_{k,\alpha\beta}t_{k,k+1}^{\alpha\beta}c_{k\alpha}^{\dagger}c_{k+1,\beta}+\text{h.c.}
\end{equation}
denotes interdot tunneling. Each dot contains a large number, $N_{0}\gg1,$ of electrons that provide a dissipative fermionic bath with an approximate
continuum of levels $\{\epsilon_{k\alpha}\}$ with mean spacing $\delta\sim O(1/N_{0}).$ The large number of single-electron levels in each dots also serves as a large-$\mathcal{N}$ parameter that provides useful simplifications leading to our final effective action (see below). In order to treat the tunneling effects correctly wrt the ``bare'' Hamiltonian which is interacting, one follows \cite{altland2010condensed} the standard process of Hubbard-Stratonovich decoupling of the interaction, and eliminating the Hubbard-Stratonovich fluctuation potentials by gauge transformations of the fermion fields. Then following Refs. \onlinecite{S_Sankar_V_Tripathi, altland2010condensed} (see also Appendix \ref{app:model}) we go over to the Keldysh path integral formalism, with the action corresponding to our Hamiltonian put on the Keldysh time contour. Thereafter the fermionic degrees of freedom are integrated out, and the resulting fermionic determinant is expanded in increasing powers of the tunneling (only even powers survive). Terms in the effective action that are $O(|t|^4)$ and higher get suppressed in the large-$\mathcal{N}$ limit.  Appendix \ref{app:model} contains an outline of these steps. The physical significance of the large-$\mathcal{N}$ approximation in suppressing higher order tunneling terms in the effective action is illustrated in Figure \ref{fig:largeN}. Note that in one or few-orbital Hubbard models, the large-$\mathcal{N}$ approximation is not available; hence, in that situation, higher order tunneling terms must be retained in the effective rotor action.

The end result is our effective Keldysh-AES rotor action,\cite{altland2010condensed,S_Sankar_V_Tripathi}
\begin{equation}
S= S_C + S_\text{tun},
\end{equation} 
for a one-dimensional array of quantum dots each
with a charging energy $E_{C}$ and interdot dimensionless conductance
$g\sim\langle |t^2|\rangle/\delta^2$,
where
\begin{align}
S_{C} & =\frac{1}{4 E_{C}}\sum_{k}\int_{t}\left[(\partial_{t}\phi_{k}^{+})^{2}-(\partial_{t}\phi_{k}^{-})^{2} + N_0 \partial_t (\phi_{k}^{+}-\phi_{k}^{-})\right],\label{eq:sc}
\end{align}
represents Coulomb correlations, and $S_{\text{tun}},$ a nonlocal term, represents the interdot tunneling
processes, 
\begin{align}
S_{\mbox{tun}}[\phi] & =g\sum_{k}\int_{t,t'}\begin{pmatrix}e^{-i\phi_{k,1}^{+}}\\
e^{-i\phi_{k,1}^{-}}
\end{pmatrix}_{t}^{T}L_{k,1}(t,t')\begin{pmatrix}e^{i\phi_{k,1}^{+}}\\
e^{i\phi_{k,1}^{-}}
\end{pmatrix}_{t'}.\label{eq:stun}
\end{align}
Here the superscripts $\pm$ respectively label the forward and backward
parts of the Keldysh time contour, the fields $\phi_{k}^{\pm}$ in
Eq. (\ref{eq:sc}) are the phases dual to the charge excitations $n_{k}^{\pm}$
(the $e^{i\phi_{k}^{\pm}}$ annihilate one charge) on the $k^{\text{th}}$
quantum dot. The phases $\phi_{k,1}^{\pm}$ in Eq. (\ref{eq:stun})
are the difference fields $\phi_{k}^{\pm}-\phi_{k+1}^{\pm}$ across
the link $(k,\,k+1).$  The
kernel $L_{k,1}(t,t')$ is a $2\times2$ matrix in Keldysh ($\pm$) space
with the following structure\cite{altland2010condensed,S_Sankar_V_Tripathi}:

\begin{equation}
L_{k,1}=\frac{1}{4}\begin{pmatrix}\Sigma_{k,1}^{R}+\Sigma_{k,1}^{A}+\Sigma_{k,1}^{K} & \Sigma_{k,1}^{R}-\Sigma_{k,1}^{A}-\Sigma_{k,1}^{K}\\
-\Sigma_{k,1}^{R}+\Sigma_{k,1}^{A}-\Sigma_{k,1}^{K} & -\Sigma_{k,1}^{R}-\Sigma_{k,1}^{A}+\Sigma_{k,1}^{K}
\end{pmatrix}_{tt'},
\end{equation}
and the functions $\Sigma^{R,A,K}$ are in turn expressed in terms
of products of the noninteracting local (in space) Green functions $G^{R,A,K},$ 
\begin{align}
\Sigma_{k,1}^{R(A)}(t,t') & =i\left(G^{R(A)}(t-t')G_{k}^{K}(t',t)\right.\nonumber \\
 & \qquad\left.+G_{k+1}^{K}(t,t')G^{A(R)}(t'-t)\right),\\
\Sigma_{k,1}^{K}(t,t') & =i\left(G_{k}^{K}(t',t)G_{k+1}^{K}(t,t')\right.\nonumber \\
 & \qquad\left.-(G^{R}-G^{A})_{t-t'}(G^{R}-G^{A})_{t'-t}\right).
\end{align}
The retarded (advanced) local Green functions have the form $G^{R(A)}=\sum_{\alpha}(i\partial_{t}\pm i\eta-\xi_{\alpha})^{-1},$
with $\xi_{\alpha}$ the $\alpha$\textsuperscript{th} single particle
energy level reckoned from the dot's Fermi level (see Appendix \ref{app:model}). The infintesimally
small positive constant, $\eta,$ ensures the theory has proper causal
structure. Likewise, $G_{k}^{K}(t,t')$ is the local Keldysh component
of the noninteracting Green function, $G_{k}^{K}(t,t')=F_{k}(t,t')(G^{R}-G^{A})_{t-t'},$
where $F_{k}$ is related to the distribution function for the single fermion excitations in a dot. 
Any power dissipated in the dots
will result in electron heating, which would necessitate tracking
the time evolution of $F_{k}.$ Therefore for simplicity, we make
the further assumption that the dot electrons are coupled to an external
phonon bath and the electron relaxation time due to electron-phonon
collisions is much shorter than that due to interdot electron tunneling,
which is of the order of $(g\delta)^{-1}.$ This allows one to replace
$L_{k,1}(t,t')$ by its \textit{equilibrium} value in which case, $L_{k,1}(t,t')$
depends only on the difference of the two time arguments \cite{S_Sankar_V_Tripathi,altland2010condensed}.
For this equilibrium case, $F_{k}(t,t')\equiv F_{k}(t-t'),$ and its
Fourier transform has the simple form $F_{k}(\epsilon)=1 - 2f_{k}(\epsilon),$
where $f_{k}(\epsilon)$ is the equilibrium Fermi distribution function.

The tunneling action in Eq. (\ref{eq:stun}) shares similarities with Josephson tunneling actions used to describe transport in superconducting dot arrays: both involve periodic functions of the phases necessary to ensure charge quantization, and also feature single-particle excitation gaps. Nevertheless, there are crucial differerences between the two. In Eq. (\ref{eq:stun}), the kernel $L_{k,1}(t,t')$ is nonlocal in time, unlike the Josephson case where it is local in time and has a form $S_\text{tun}[\phi]=J \sum_k \int_t \cos(2\phi(t)).$ The time non-locality arises from integrating out the gapless particle-hole fermionic excitations in the origin and destination dots and represents the dissipative nature of the interdot tunneling process. In a JJ array, the tunneling of Cooper pairs occurs without dissipation, and particle-hole excitations are subject to the superconducting gap $\Delta.$ Another difference is that the Josephson coupling $J\sim g\Delta$ explicitly depends on the superconducting gap whereas for the normal case, the tunneling term has no characteristic energy scale. However both superconductor and normal dot chains do contain the charging term that has a characteristic scale $E_C.$ We shall see below that the similarities of the two cases result in similar optical response, and serves to caution relying on certain optical properties for confirming superconductivity. However the differences between the two cases show up in properties such as the absence of Shapiro steps - a key signature of supercurrent - in the normal dot array.

The retarded(advanced) Green function have a causal structure, i.e.,
$G^{R}(t)\propto\Theta(t)$ and $G^{A}(t)\propto\Theta(-t)$. The
same causality structure is obeyed by $\Sigma^{R(A)}(t)$. Additionally,
in Fourier space the following identities facilitate the calculation
of expectation value of current response:
\begin{align}
\left(\Sigma_{k,1}^{R}-\Sigma_{k,1}^{A}\right)_{\epsilon}= & \dfrac{i}{\pi}\epsilon,\\
\left(\Sigma_{k,1}^{K}\right)_{\epsilon}= & \dfrac{i}{\pi}\epsilon F_b(\epsilon),
\end{align}
where $F_b(\epsilon) = 1 + 2f_b(\epsilon),$ with $f_b$ the equilibrium Bose distribution (see Appendix \ref{app:model}). In the rest of this paper, we choose to work in units $\hbar=e=1.$

We introduce the external AC electric field as a time-dependent ``classical'' vector potential on the links,
$A_{k,1}^{c}=(A_{k,1}^{+}+A_{k,1}^{-})/2,$ turned on at time
$t=0$ (a quench),
\begin{align}
A_{k,1}^{c} & =\Theta(t)(V/\Omega)\cos(\Omega t).\label{eq:Acl}
\end{align}
 This changes the tunneling part of the action, Eq. (\ref{eq:stun}), by incorporating Peierls shifts in the phase differences, $\phi_{k,1}^{c,q}(t)\rightarrow\phi_{k,1}^{c,q}(t)+A_{k,1}^{c,q}(t),$ where $\phi_{k,1}^{\pm}= \phi_{k,1}^{c}\pm\phi_{k,1}^{q}/2.$ The
nonequilibrium current, $\hat{j}_{k,1}[\phi],$ in a link is obtained
in the usual manner by introducing infinitesimal quantum components
of the vector potential, $A_{k,1}^{q},$ and varying the action with
respect to it. 

Defining  $e_{k,1}^{\pm}(t)= \exp(i \phi_{k,1}^{\pm}(t)-i A_{k,1}^{\pm}(t)) $	,
the current functional in terms of phase fields is
\begin{align}
\hat{j}_{k,1}(\tau) & =ig\int_{t}\Bigl[(e_{\tau}^{+})^{*}L_{\tau,t}^{++}e_{t}^{+}-(e_{t}^{+})^{*}L_{t,\tau}^{++}e_{\tau}^{+}\nonumber \\
 & +(e_{\tau}^{+})^{*}L_{\tau,t}^{+-}e_{t}^{-}+(e_{t}^{+})^{*}L_{t,\tau}^{+-}e_{\tau}^{-}-(e_{\tau}^{-})^{*}L_{\tau,t}^{-+}e_{t}^{+}\nonumber \\
 & -(e_{t}^{-})^{*}L_{t,\tau}^{-+}e_{\tau}^{+}-(e_{\tau}^{-})^{*}L_{\tau,t}^{--}e_{t}^{-}+(e_{t}^{-})^{*}L_{t,\tau}^{--}e_{\tau}^{-}\Bigl],
 \label{eq:current-exp}
\end{align}
where we have skipped the site indices for brevity. The above expression for the current response is formally exact; however the averaging over the fields gives contributions in increasing powers of the tunneling conductance, with the leading order corresponding to the atomic limit.

\section{Nonequilibrium AC current response}\label{sec:ACresponse}

We are interested in the current response of our infinite array deep in the Mott insulator
regime, $g\ll1,$ where transport is dominated by $O(g)$ sequential tunneling terms since at any order in tunneling, the cost $2E_C$ for creating a particle-hole pair is always paid. In contrast, higher order tunneling effects are important in short arrays, or where $g \gtrsim 1.$ Interdot tunneling processes also lead to the renormalization of the charging energy $E_{C}$ due to virtual
tunneling to nearby dots. It is known that such processes result in $O(g)$ corrections to $E_{C},$ which for $g\ll1$ do not make a
qualitative difference to our findings except for replacing $E_{C}$
by its renormalized value \cite{Falci}. In contrast, when $g\sim O(1),$ the effect of the surrounding dots becomes important, and indeed, Coulomb blockade effects tend to get washed out, and the current response becomes resistive \cite{RevModPhys_aes2}.

\begin{figure}
\includegraphics[width=1\columnwidth]{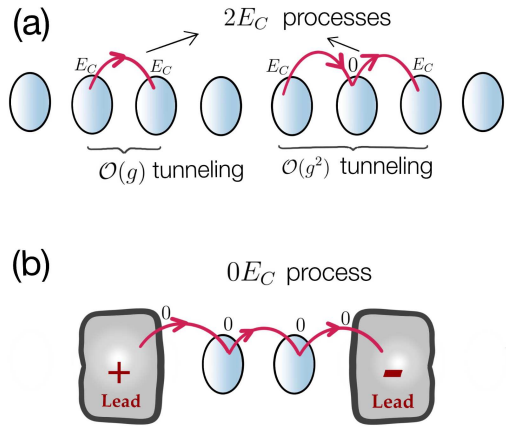}
\caption{\label{fig:cotunn}(a) Illustration of the dominance of single-link tunneling processes in transport in the infinite chain deep in the Mott insulating regime. Since a Coulomb blockade cost has to be always paid for creating a particle-hole pair at different sites, the single link tunneling process that is $O(g)$ dominates higher order processes. In (b), a system with a small number of dots is shown. Here a higher order cotunneling process from source to drainn leads allows one to circumvent Coulomb blockade at the dots since only virtual excitations occur in the intermediate dots. Note that the leads by definition have zero charging energy.}
\end{figure}

In the rest of our analysis, we shall compute only the leading order contribution to the current response. Figure \ref{fig:cotunn}  provides a pictorial illustration of why higher order tunneling effects are not significant in infinite quantum dot arrays deep in the Mott insulator regime.

Instead of working with the charging action of Eq. (\ref{eq:sc}) that involves only the phases, it is convenient to work with a number-phase respresentation. For this, we rewrite the action in Eq. (\ref{eq:sc}) as $\int_t [1/2E_C(\partial_t \phi_{k}^{c})(\partial_t \phi_{k}^{q})+N_0\partial_t\phi_{k}^{q}],$ and decouple the first term introducing Hubbard-Stratonovich fields $n_{k}^{q,c}:$ 
\begin{equation}
S_C[n,\phi]=\sum_{k}\int_{t}\left([n_{k}^{c}+N_{0}]\partial_{t}\phi_{k}^{q}+n_{k}^{q}\partial_{t}\phi_{k}^{c}-2E_{C}n_{k}^{c}n_{k}^{q}\right).\label{eq:number-phase}
\end{equation}
Here $n_{k}^{c(q)}$ is the classical (quantum) component of the charge
excitation on the $k^{\text{th}}$ dot and is conjugate to $\phi_k^{q(c)}.$ At this stage, the charge variable $n$ is unconstrained. We allow for finite winding numbers in $\phi_{k}^{q},$ 
\begin{equation}
 \phi_{k}^{q}(t) \rightarrow \frac{2\pi W_k}{T_0}(t-T_0)+ \phi_{k}^{q}(t),
\end{equation}
where $T_0$ is a time scale longer than any others in the problem, and the phase on the RHS satisfies Dirichlet boundary conditions, $\phi_{k}^{q}(0)=\phi_{k}^{q}(T_0)=0.$ Summing over the winding numbers forces $n_{k}^{c}$ to take only integer values. An in-depth discussion of the charge quantization emerging from such considerations can be found in the literature, for example in the textbook Ref. \onlinecite{altland2010condensed}.

Performing the average with respect to the phase fields, $\int D\phi\hat{j}_{k,1}[\phi]e^{iS},$
yields the expectation value $j(t)$ for the current. To leading order in the perturbation series, 
the averaging requires calculation of bare bond correlators defined as 
\begin{equation}
\Pi_{\sigma\sigma'}^{(0)}(t,t')=\left\langle \exp[-i\phi_{k,1}^{\sigma}(t)+i\phi_{k,1}^{\sigma'}(t')]\right\rangle _{0},
\end{equation}
where $\left\langle ...\right\rangle _{0}$ denotes averaging with
the bare action. The bare bond correlator can be factorised into a
product of two single site correlators:
\begin{equation}
\Pi_{\sigma\sigma'}^{(0)}(t,t')=C_{\sigma\sigma'}(t,t')C_{\sigma'\sigma}(t',t),
\end{equation}
where 
\begin{equation}
C_{\sigma\sigma'}(t,t')=\left\langle \exp[-i\phi_{k}^{\sigma}(t)+i\phi_{k}^{\sigma'}(t')]\right\rangle _{0},
\end{equation}
and the site correlators can be shown to be\cite{S_Sankar_V_Tripathi}
\begin{align}
C_{\pm\pm}(t,t')= & \exp[\mp iE_{C}|t-t'|],\\
C_{\pm\mp}(t,t')= & \exp[\pm i E_{C}(t-t')].
\end{align}

The expectation value of the current response so obtained is 

\begin{widetext}

\begin{gather}
j(\tau)=ig\int_{-\infty}^{\tau}e^{iA_{k,1}^{c}(\tau)-iA_{k,1}^{c}(t)}\Bigl\{\Sigma_{(\tau-t)}^{R}\cos[2E_{C}(\tau-t)]
-i\Sigma_{(\tau-t)}^{K}\sin[2E_{C}(\tau-t)]\Bigl\} dt+\text{c.c.}
\end{gather}
We use $A^c_{k,1}(t) =  \frac{V}{\Omega} \cos(\Omega t)$, and
expand the exponentials containing the vector potential making
use of the Jacobi-Anger formula, $e^{iz\cos(x)}=\sum_{n=-\infty}^{\infty}i^{n}J_{n}(z)e^{inx},$
and perform the time integrals to obtain our final expression for $j(\tau)$ following the quench:

\begin{align}
j(t) & =-\frac{4gE_{C}}{\pi^{2}}\left[\sin\left(\frac{V}{\Omega}\cos(\Omega t)\right)\left\{ \text{Ci}(2E_{C}t)\left(1-J_{0}\left(\frac{V}{\Omega}\right)\right)-\text{sinc}(2E_{C}t)\right\} -\sin\left(\frac{V}{\Omega}[1-\cos(\Omega t)]\right)\text{sinc}(2E_{C}t)\right]\nonumber \\
 & -\frac{4gE_{C}}{\pi^{2}}\sin\left(\frac{V}{\Omega}\cos(\Omega t)\right)J_{0}\left(\frac{V}{\Omega}\right)\ln\left(\frac{2E_{C}}{\Omega}\right)\nonumber \\
 & -\frac{2g\Omega}{\pi^{2}}\sum_{n=1}^{\infty}(-1)^{n}J_{n}\left(\frac{V}{\Omega}\right)\sin\left(\frac{V}{\Omega}\cos(\Omega t)+\frac{n\pi}{2}\right)\left[\left(\frac{2E_{C}}{\Omega}-n\right)I_{n}(t)+\left(\frac{2E_{C}}{\Omega}+n\right)I_{-n}(t)\right.\nonumber \\
 & \left.+\cos(n\Omega t)\left\{ \left(\frac{2E_{C}}{\Omega}-n\right)\ln\left|\frac{2E_{C}}{\Omega}-n\right|+\left(\frac{2E_{C}}{\Omega}+n\right)\ln\left(\frac{2E_{C}}{\Omega}+n\right)\right\} \right].\label{eq:current}
\end{align}
\end{widetext}
Here the $J_{n}$ are Bessel functions of the first
kind, $\text{sinc}(y)=\sin y/y,$ $I_{n}(t)=-\text{Ci}[(2E_{C}-n\Omega)t]\cos(n\Omega t)+\text{si}[(2E_{C}-n\Omega)t]\sin(n\Omega t),$
and the functions $\text{Ci}$ and $\text{si}$ are respectively the
trigonometric integrals, $\text{Ci}(y)=-\int_{y}^{\infty}dx\cos x/x$
and $\text{si}(y)=-\int_{y}^{\infty}dx\sin x/x.$
We have used here the standard notation for the trigonometric integrals, the sine counterpart, $\text{Si}(y)=\int_{0}^{y}dx\,\sin(x)/x,$ of $\text{Ci}(y)$ is related to $\text{si}(y)$ through $\text{Si}(y)=(\pi/2)+\text{si}(y).$

The physical significance of the different contributions to the current
in Eq. (\ref{eq:current}) may be understood as follows. The first
line in Eq. (\ref{eq:current}) describes Bloch-like oscillations
at a frequency $2E_{C}.$ The amplitude of the Bloch-like oscillations,
a signature of charge quantization, falls inversely with time following
the quench owing to the presence of dissipation. Such decay of Bloch oscillations
is not seen in KDMFT studies of dissipationless half-filled Hubbard
chains \cite{P_Werner_Floquet_dc_ac}, and could be a consequence of insufficiently long waiting time in 
the numerical simulations.

At long times, only the contributions from the last three lines of
Eq. (\ref{eq:current}) survive and the NESS current is obtained by
simply making the substitution, 
\begin{align*}
\left(2E_{C}/\Omega-n\right)I_{n}(t)+\left(2E_{C}/\Omega+n\right)I_{-n}(t)\\
 & \!\!\!\!\!\!\!\!\!\!\!\!\!\!\!\!\!\!\!\!\!\!\!\!\!\!\!\!\!\!\!\!\!\!\!\!\!\!\!\!\!\!\!\!\!\!\!\!\!\!\!\!\!\!\!\!\!\!\!\!\!\!\!\!\!\!\!\!\!\!\!\!\!\!\!\!\!\!\!\!\rightarrow-\pi(n-2E_{C}/\Omega)\Theta(n-2E_{C}/\Omega)\sin(n\Omega\tau).
\end{align*}
The parameter $2E_{C}/\Omega$ is the number $n$ of photons required
to excite an electron through the Mott gap, while $V/\Omega$ controls
the strength, $J_{n}(V/\Omega),$ of an $n$-photon process. The logarithmic singularities at the thresholds $n\Omega=2E_C$ are not the expected response of a gapped system but reflect the collective response of the dot electrons upon a tunneling event, similar to the X-ray edge phenomenon. The similarity
with X-ray edge singularity is there because each dot has a large
number of electrons and every tunneling event shifts the entire electron
band by a large amount, $E_{C}.$ 

\begin{figure*}
\includegraphics[width=1\textwidth]{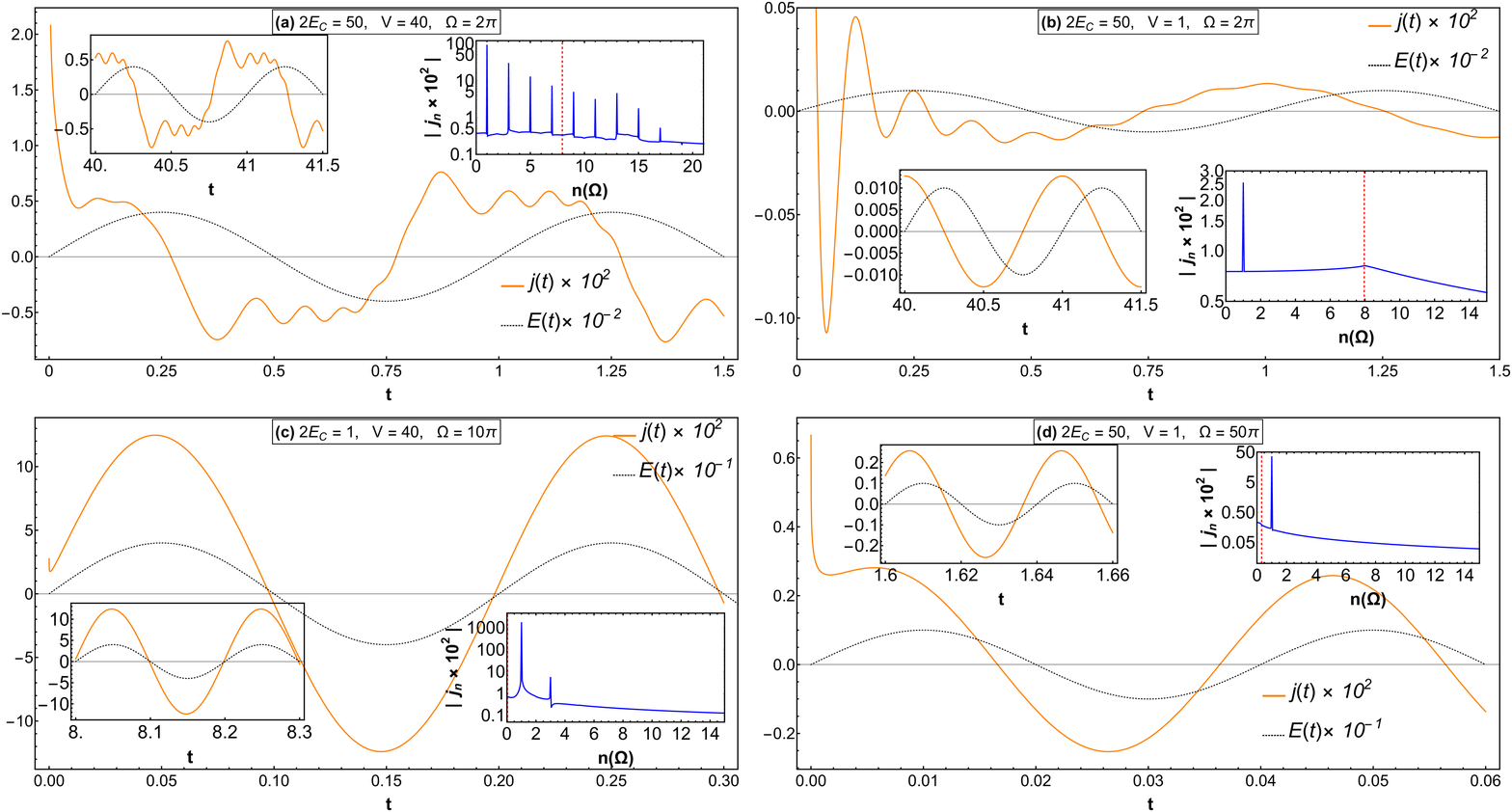}

\caption{\label{fig:four_cases}Plots showing the transient current response
$j$ (solid curves) vs time $t$ following an AC quench
(dotted curves), in different regimes of the parameters $2E_{C}/\Omega$
and $V/\Omega.$ The left insets in (a)-(d) show the late time current
response (solid curves) together with the AC field (dotted curves),
and the right insets show the power spectrum of the current response
as a function of frequency, in units of the driving field frequency,
$\Omega.$ The dashed vertical lines in the power spectrum plots indicate
the position of $2E_{C}$ corresponding to the Bloch-like oscillations
(visible only in (b) where AC induced charge fluctuations are weak).
Peaks at odd multiples of $\Omega$ are seen in the power spectra
when $V/\Omega\gg1,$ and correspond to multiphoton assisted tunneling
processes. In (c), the long time current response is in phase with
the drive, like a resistor, while in the other cases, there is a finite
phase difference.}
\end{figure*}

Figure \ref{fig:four_cases} shows the transient current response $j(t)$
following the AC quench, in four different parameter regimes. 
The left insets in (a)-(d)
show the late time current response, together with the AC field, while
the right insets show the power spectrum of $j(t)$ as a function
of frequency, measured in units of $\Omega.$ The power spectra all
show a peak at $\Omega,$ and for stronger field strengths, $V/\Omega>1,$
higher harmonics appear at odd integer multiples of $\Omega$ corresponding
to multiphoton processes.
The vertical dashed lines in
the power spectra indicate the position of $2E_{C}$ at which the
Bloch-like oscillations, a signature of charge quantization, occur. 

Figures \ref{fig:four_cases}(a), (b) correspond to $2E_{C}/\Omega\gg1,$
where the Mott gap greatly exceeds the driving frequency. In both
cases the phase of the current response is approximately $\pi/2$
ahead of the driving field, which, in combination with the insulating
frequency dependence (see analysis below), resembles a capacitor.
The power spectrum of the current in Fig. \ref{fig:four_cases}(a),
where $V/\Omega>1,$ shows significant multiphoton peaks, but a distinct
signature of the Bloch-like oscillations at $2E_{C}$ is not evident.
In contrast, the Bloch-like peak at $2E_{C}$ is clearly visible in
Fig. \ref{fig:four_cases}(b) where $V/\Omega<1,$ so that multiphoton
processes that could excite electrons across the Mott gap, are suppressed.
Since power dissipation is governed by the component of the current
that is in phase with the driving field, both (a) and (b) correspond
to weak dissipation. In Fig. \ref{fig:four_cases}(c), where $2E_{C}/\Omega\ll1$
and $V/\Omega>1,$ charge excitations induced by both single and multiphoton
processes are significant. The current response is predominantly at
the driving frequency, and in phase with the AC field, much like a
resistor, which one would expect \cite{oka_bethe_ansatz} due to pair
production facilitated by the small Mott gap and large electric field
strength. In Fig. \ref{fig:four_cases}(d), the large driving frequency
implies single-photon dominated charge excitations. We
show below that the frequency dependence is that
of a resistively shunted inductor. Similar odd harmonic generation
and multiphoton assisted tunneling phenomena have also been reported
in recent KDMFT-based numerical studies \cite{P_Werner_HHG,P_Werner_Floquet_dc_ac,oka_philipp_eckstein_Dielectric_Breakdown}.

\begin{figure*}
\includegraphics[width=0.95\textwidth]{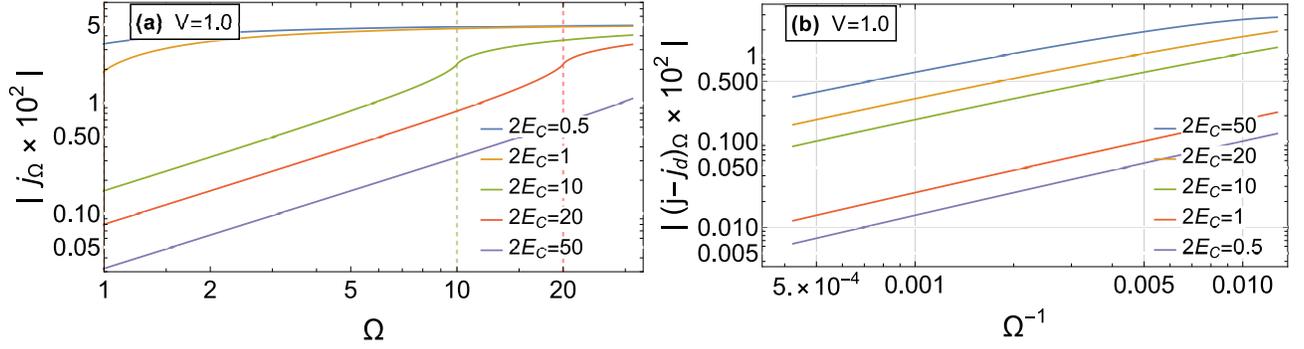}

\caption{\label{fig:Jomega}Plots showing the frequency dependence of the single-photon 
component $j_{\Omega}$ of the current for a given AC electric field, $V=1.$ 
In (a), a capacitative Mott insulator type
response, $|j_{\Omega}|\sim\Omega,$ is seen at low frequencies ($2E_{C}/\Omega\gg1$)
and weak electric fields ($2E_{C}/V\gg1$). 
Singularities in $j_{\Omega}$ appear 
at $\Omega=2E_{C}$ separating insulator and conductor
like behavior of the current. 
In (b), the behavior of $j_{\Omega}$ at high frequencies and low fields, $2E_{C}/\Omega,\,V/\Omega\ll1,$
is like a resistively shunted inductor, i.e., $|(j-j_{d})_{\Omega}|\sim\Omega^{-1},$
where $j_{d}$ is the dissipative component of the current. }
\end{figure*}

We now examine the limits where the current response is capacitative
or inductive. For simplicity we consider the single-photon dominated
regimes where $V/\Omega\ll1,$ and Eq. (\ref{eq:current}) for the
NESS current simplifies to ($x=2E_{C}/\Omega$)
\begin{align}
j & \approx V\frac{2g}{\pi^{2}}\begin{cases}
\frac{1}{x}\cos(\Omega t), & x\gg1,\\
\frac{\pi}{2}\sin(\Omega t)+x\ln(\frac{1}{x})\cos(\Omega t), & x\ll1.
\end{cases}\label{eq:NESS}
\end{align}
At low frequencies where $x\gg1,$ the current is proportional to
the derivative of the driving field - a capacitative response characteristic
of a Mott insulator. At high frequencies, $x\ll1,$ single-photon
processes are sufficient to ensure charge excitations across the Mott
gap, and the current is approximately linear-response type, having
components proportional to the driving field as well as to its time
integral. Apart from an additional enhancement by a factor $\ln(\Omega/2E_{C}),$
it is essentially that of a resistively shunted inductor, with \emph{negative}
link inductance $L\sim-[gE_{C}\ln(\Omega/2E_{C})]^{-1}$. While this behavior also superficially resembles that of a classical $RC$ series circuit; however, the $1/\Omega$ term there has a very different dependence on $gE_C,$ which is the interdot Thouless energy of diffusion of a particle-hole pair. This parameter enters our expression in a manner similar to the Josephson energy in superconductor dot arrays.
The inductive response is also unrelated to the surface
plasmon related Mie resonance that occurs in the same system \cite{tripathi2006frequency}.
As $x\rightarrow0,$ the current response is resistive,
and independent of $E_{C}.$ Figure \ref{fig:Jomega} shows the magnitude
of the dominant single-photon component $j_{\Omega}$ of the current
in both these frequency limits. Kinks in $j_{\Omega}$ curves are
seen at $\Omega=2E_{C}.$ The logarithmic non-analyticities separating insulating and metallic frequency dependence of the linear response current indicate an insulator to metal transition. 

\begin{figure}
\includegraphics[width=1\columnwidth]{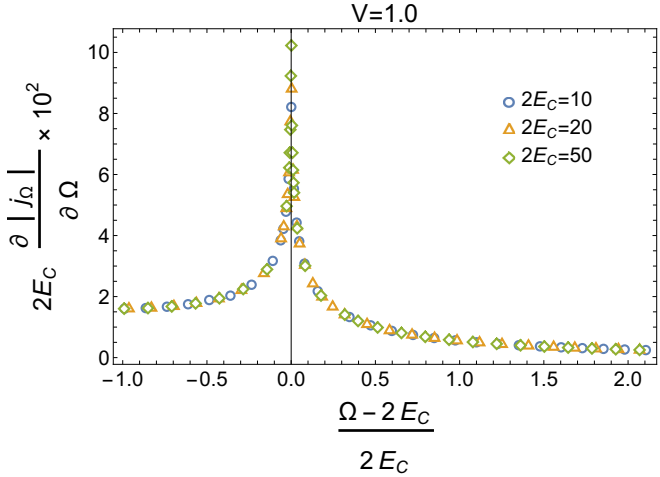}
\caption{\label{fig:scaling}Plot illustrating the behavior of $\partial|j_{\Omega}|/\partial\Omega$
in the presence of a purely AC drive for different values of $E_{C},$
and a fixed value of $V=1.$ By rescaling the axes, the curves are
seen to collapse, indicating a universal response that depends only
on the dimensionaless frequency $\Omega/2E_{C}.$ The analytic dependence
is given by Eq. \ref{eq:current}.}
\end{figure}

Figure \ref{fig:scaling} shows the behavior of the frequency derivative, $\partial|j_{\Omega}|/\partial\Omega,$ of the single photon component of the current response as a function of the distance from the threshold $\Omega=2E_C.$ This provides a clearer illustration of the frequency driven insulator to metal transition. Upon crossing
the threshold $\Omega=2E_{C},$ the current response changes from
insulating (capacitative) to metallic, which is seen in the increase
and subsequent decrease of $\partial|j_{\Omega}|/\partial\Omega$
through the threshold. It is evident from the expression for the current
response, Eq. \ref{eq:current} that the singularity in the current
response at $\Omega=2E_{C}$ is essentially logarithmic. The collapse of the curves for
different parameter values indicates a universal singular response that depends only on the 
dimensionless frequency $\Omega/2E_{C}.$

\begin{figure*}
\includegraphics[width=0.9\columnwidth]{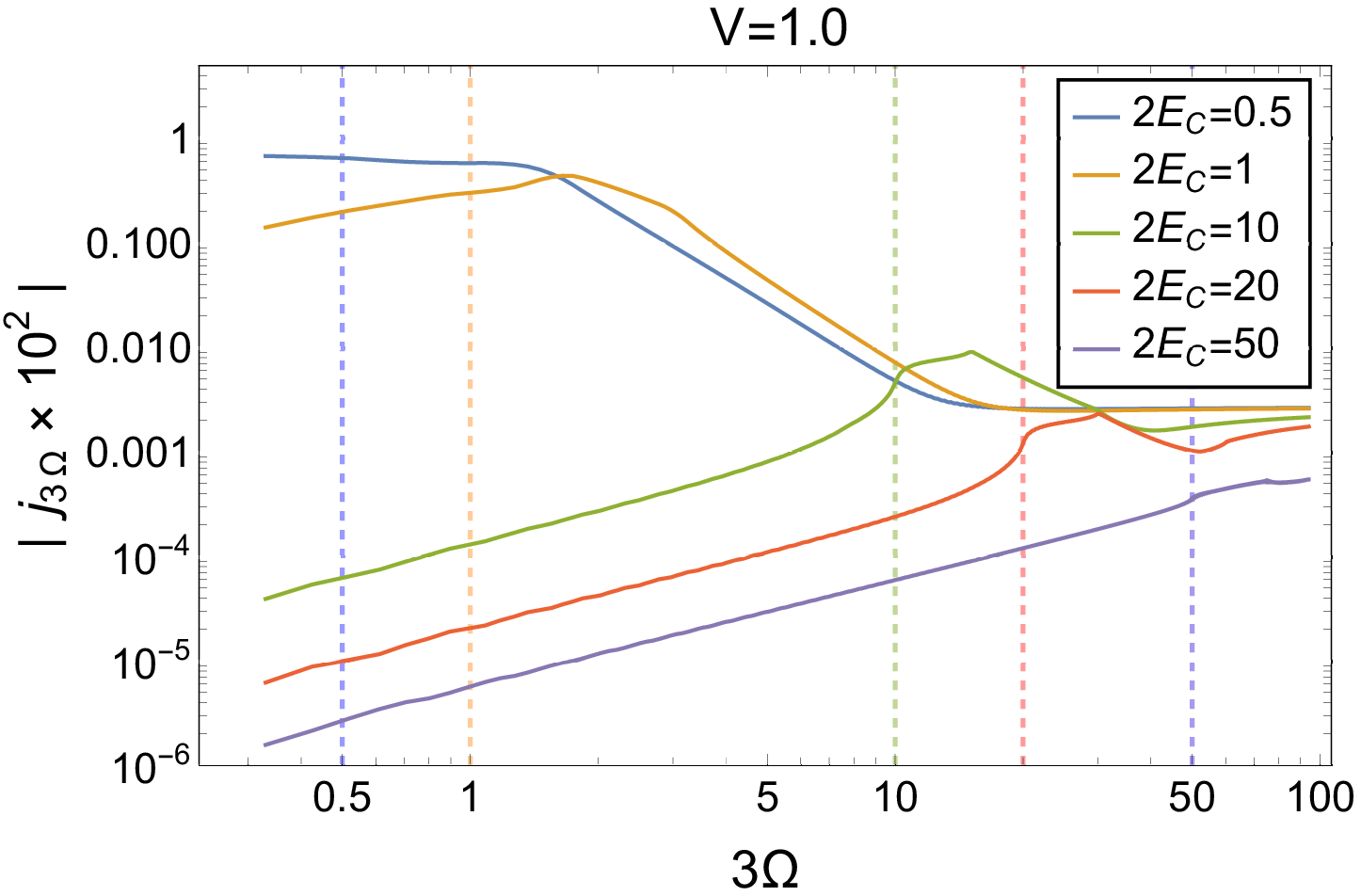}\qquad{}\vspace{20bp}
\includegraphics[width=0.9\columnwidth]{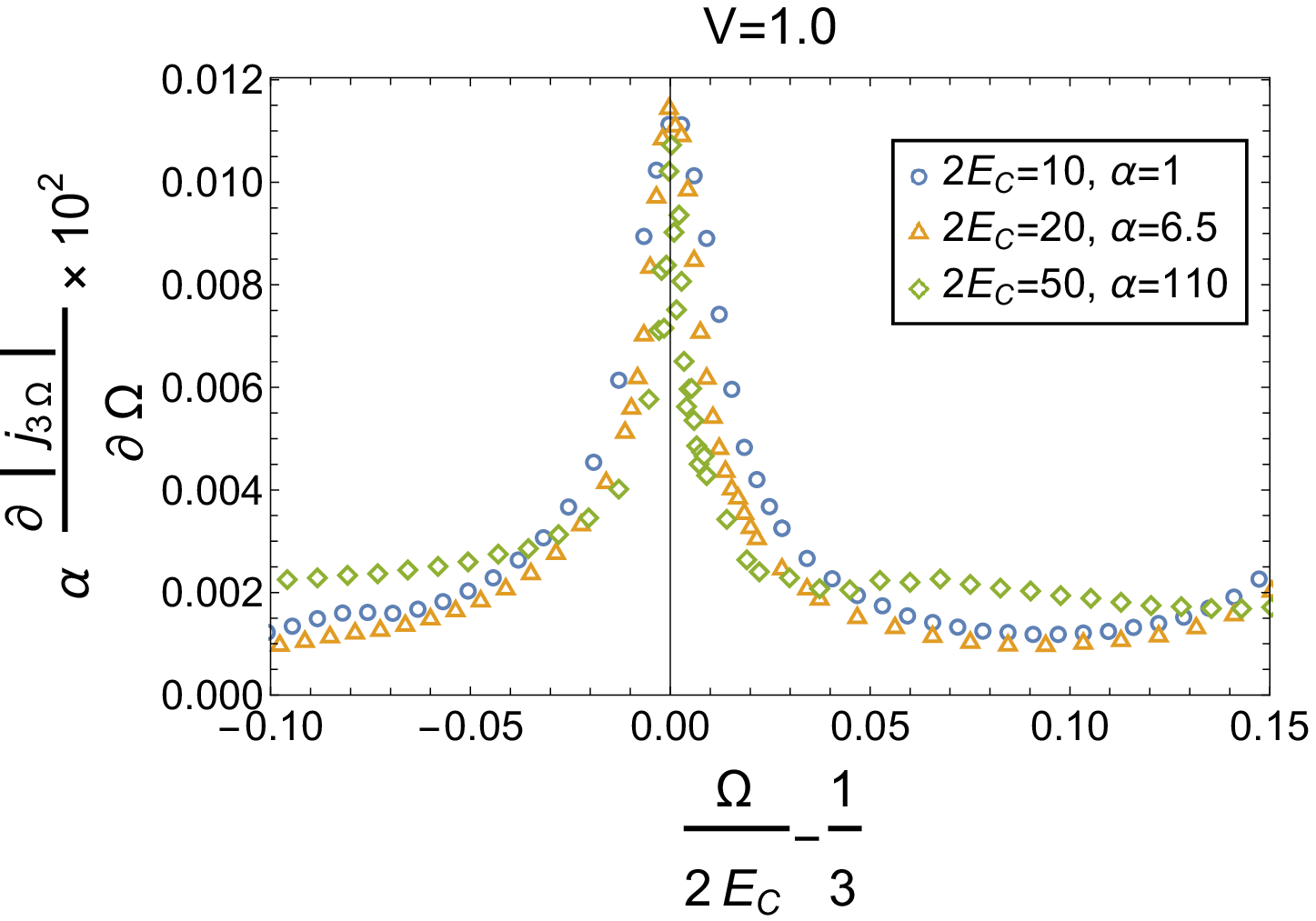}

\caption{\label{fig:j3}The plot on the left shows the current response $j_{3\Omega}$
corresponding to the third harmonic, to be viewed in comparison with
the linear response component $j_{\Omega}.$ Kink-like features are
seen when the threshold condition $3\Omega=2E_{C}$ for three-photon
excitation across the Mott gap is satisfied. Note that $j_{3\Omega}$
is much smaller than the linear response component. Right plot shows
the derivative of $j_{3\Omega}$ in the vicinity of the threshold.
The singularity in the derivative of the current response is on account
of the logarithmic singularities in the current
response. The data collapse shows universality in the current response
for different values of the Coulomb energy.}
\end{figure*}

Multiphoton processes become important at low frequencies, $V/\Omega\gg 1.$ This is evident in Fig. \ref{fig:four_cases}a. We found that these processes are typically much smaller than the single photon contribution. In Fig. \ref{fig:j3} we present plots for the third harmonic (and the derivative) of the current response. The weak logarithmic
singularities are present at the threshold $3\Omega=2E_{C},$ and
there is a scaling collapse at the threshold, similar to the single
photon case. The three-photon contribution is smaller than the dominant
single-photon contribution by a factor of around 100.

This NESS response exhibits remarkable
similarities with resistively shunted Josephson arrays. The high frequency regime
resembles that of a superconducting Josephson
array albeit with \emph{negative} stiffness, $K=-gE_{C}\ln(\Omega/2E_{C}),$ amounting to $\pi$-phase slips
in the links like an $\eta$-pairing phase \cite{yang_eta_pairing_hubbard_chain}. This should be compared to the Ambegaokar-Baratoff relation, $K=+g\Delta$
for the stiffness of the Josephson junction, where $\Delta$ is the
gap to quasiparticle excitations in the dots.
This effect is reminiscent of the reported AC induced attractive electron interactions enhancement
reported in studies of the optically excited half-filled Hubbard models
 \cite{Olivier_ac_driven,Tomonori_Shirakawa_photoinduced_eta_pairing,nature_pairing_due_to_ac,clark_enhancementOfpairing_ac,philipp_eckstein_long_range_eta_pairing,PRL_Metastable_superfluidity_Rosch,science_kota,werner_eckstein_entropy_doublonHolon,T_Oka_P_Werner_ac_Repulsive_to_Attractive}. Various
groups have subsequently also reported the formation of an $\eta$-pairing
state in the presence of significant doublon excitations \cite{Millis2019optically,PRL_Tindall_long_eta_pairing,Tomonori_Shirakawa_photoinduced_eta_pairing}.
Note however that in our case, although the resemblance with $\eta$-pairing
is there, it cannot be attributed to doublon production and condensation
since the equilibrium electron number in the quantum dots is arbitrary. The question of existence of supercurrent 
can be addressed by looking for Shapiro steps in the current response under simultaneous AC and DC bias.

\section{NESS for simultaneous AC and DC bias}\label{sec:shapiro}
To probe whether there is indeed an AC induced
superconducting $\eta$-pairing state in our case, we analyze below
the current response in the presence of a simultaneous DC field of strength $D,$ for which we choose 
\begin{equation}
A^c_{k,1}(t) =  \frac{V}{\Omega} \cos(\Omega t) + D t.
\end{equation}
The NESS current response for this case may be obtained following
the same procedure outlined above for the AC response. 
The
expression for $j(t),$ although straightforward to obtain, is rather
cumbersome, and we introduce the following additional quantitites
to simplify its presentation:
\begin{align*}
p_{1} & =\pi\Theta(n\Omega-2E_{C}-D)(n\Omega-2E_{C}-D),\\
p_{2} & =\pi\Theta(-n\Omega-2E_{C}-D)(-n\Omega-2E_{C}-D),\\
p_{3} & =\pi\Theta(n\Omega-2E_{C}+D)(n\Omega-2E_{C}+D),\\
p_{4} & =\pi\Theta(-n\Omega-2E_{C}+D)(-n\Omega-2E_{C}+D),
\end{align*}
and
\begin{align*}
q_{1} & =(n\Omega-2E_{C}-D)\ln|n\Omega-2E_{C}-D|,\\
q_{2} & =(-n\Omega-2E_{C}-D)\ln|-n\Omega-2E_{C}-D|,\\
q_{3} & =(n\Omega-2E_{C}+D)\ln|n\Omega-2E_{C}+D|,\\
q_{4} & =(-n\Omega-2E_{C}+D)\ln|-n\Omega-2E_{C}+D|,
\end{align*}
 and $\theta_{\pm}=\dfrac{V}{\Omega}\cos(\Omega t)\pm n\Omega t$.

Now, introducing the quantities $f_{n}$ and $g_{n},$
\begin{gather*}
f_{n}=p_{1}\cos\theta_{+}+p_{2}\cos\theta_{-}-p_{3}\cos\theta_{-}-p_{4}\cos\theta_{+}\\
\qquad-q_{1}\sin\theta_{+}-q_{2}\sin\theta_{-}-q_{3}\sin\theta_{-}-q_{4}\sin\theta_{+},
\end{gather*}
 
\begin{gather*}
g_{n}=-p_{1}\sin\theta_{+}-p_{2}\sin\theta_{-}+p_{3}\sin\theta_{-}+p_{4}\sin\theta_{+}\\
\qquad-q_{1}\cos\theta_{+}-q_{2}\cos\theta_{-}-q_{3}\cos\theta_{-}-q_{4}\cos\theta_{+},
\end{gather*}
we present our final expression for the current response in the presence of a simultaneous DC and AC bias:

\begin{widetext}

\begin{gather}
j(t)=\dfrac{g}{2\pi^{2}}J_{0}\left(\dfrac{V}{\Omega}\right)f_{0}-\dfrac{g}{\pi^{2}}\mathop{\sum_{{\scriptscriptstyle {\scriptstyle n=4k,k=0}}}^{{\scriptscriptstyle {\scriptstyle k=\infty}}}\biggl\{ J_{n}\left(\dfrac{V}{\Omega}\right)f_{n}}-J_{n+1}\left(\dfrac{V}{\Omega}\right)g_{n+1}-J_{n+2}\left(\dfrac{V}{\Omega}\right)f_{n+2}+J_{n+3}\left(\dfrac{V}{\Omega}\right)g_{n+3}\biggl\}.\label{eq:SI-JDCAC}
\end{gather}

\end{widetext}

Here, $J_{n}(x)$ are Bessel's function of the first kind. To get
the DC component, $j_{DC}$ we numerically average the current over
a large time interval. The DC transconductance is $\frac{\partial j_{DC}}{\partial D}$. 

\begin{figure}
\includegraphics[width=0.95\columnwidth]{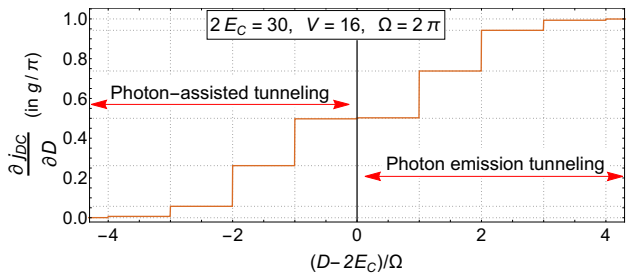}
\caption{\label{fig:Shapiro}Plot showing photon assisted tunneling and stimulated
photon emission phenomena in the presence of a simultaneous AC drive
$V\sin(\Omega t)$ and DC bias $D>0.$ The $y$-axis shows the leading
contribution to DC transconductance $\partial j_{DC}/\partial D$
as a function of the DC bias. For $D<2E_{C},$
the steps correspond to tunneling assisted by absorption of one or
more photons, and for $D>2E_{C},$ to tunneling
with stimulated photon emission. The threshold for the finite DC current
is at $D/\Omega=[2E_{C}/\Omega]+1.$ At large DC bias, the DC
conductance approaches $g/\pi$ signifying strong charge fluctuations
and absence of Coulomb blockade. Comparing with driven Josephson junctions,
$E_{C}$ plays the role of the single particle gap, $\Delta.$}
\end{figure}

We observe jumps in the DC transconductance, $\partial j_{DC}/\partial D,$
at specific values of the DC potential difference across a link $D=D_{n}:$
\begin{align}
|D_{n}|- & 2E_{C}=n\Omega,\label{eq:shapiro}
\end{align}
where $n$ is an integer, like the photon-mediated tunneling steps
in the $IV$ characteristics of resistively shunted Josephson arrays
\cite{tinkham2004introduction}. The role of the quasiparticle gap,
$\Delta,$ in the Josephson case is taken by the Coulomb scale $E_{C}$
here. However, in contrast with the Josephson case, additional Shapiro
steps that appear in intervals of $\Omega/2$ and are associated with
the supercurrent, are absent here. For $n>0,$ DC transport occurs
through stimulated emission of photons while for $n<0,$ photon-assisted
tunneling takes place. Figure \ref{fig:Shapiro} shows the DC transconductance
when a simultaneous DC bias, $D,$ is also applied. In the absence
of the AC field, the DC transconductance is known to have a threshold
behavior \cite{matsuura2008CDW_shapiro}, vanishing for $|D|<2E_{C},$
and assuming the value $g/\pi$ for $|D|>2E_{C}.$ When the AC drive
is also turned on, the threshold shifts to a lower value and transconductance
is zero for $|D|/\Omega<[2E_{C}/\Omega]+1,$where $[a]$ is the greatest
integer less than or equal to $a.$ This process is photon-assisted
tunneling. When $|D|>2E_{C},$ steps continue to appear in the transconductance
but now they are associated with tunneling accompanied by stimulated
photon emission.

\section{Discussion}\label{sec:discussion}
To summarize, we showed that the current response to an AC drive in
our dissipative Mott insulator chain undergoes a transition
from an insulator frequency dependence at low frequencies $\Omega/2E_{C}\ll1$ to
a conducting, diamagnetic frequency dependence for high frequencies $\Omega/2E_{C}\gg1,$
with an effective phase stiffness $K \sim gE_C,$ the Thouless energy for interdot diffusion of a particle-hole pair. The transition occurs at the threshold $\Omega = 2E_C$ for photon-assisted transition across the Mott gap. The presence of a large number of electrons in the dots results in characteristic logarithmic singularities reminiscent of X-ray edge phenomena in atomic physics. At high frequencies the sign of the diamagnetic response
is negative and resembles $\eta$-pairing \cite{yang_eta_pairing_hubbard_chain}
in the half-filled Hubbard chain. We argued that the $\eta$-pairing
like behavior in our model is not due to superconductivity but a consequence
of strong charge fluctuations brought about by the nonequilibrium
drive. This view is confirmed by the absence of Shapiro steps, a key
signature of supercurrent, and also the fact that despite the diamagnetic behavior at
high frequencies, the charge stiffness is zero.

For our analysis, we employ an analytical Keldysh field theory approach based on the Ambegaokar-Eckern-Sch\"{o}n rotor model for studying electronic transport in quantum dot arrays. Our technique correctly reproduces a number of known results on optically driven Hubbard chains (e.g. Bloch-like oscillations, odd harmonic generation, apparently attractive Coulomb correlations) obtained in numerical Keldysh-DMFT studies of optically excited half-filled Hubbard chains. However, in contrast with the numerical studies, we find a slow power-law decay of the Bloch-like oscillations. This could have been missed in earlier DMFT studies due to insufficient time lapse in the simulations following a quench. 

Our treatment suggests that caution must be exercised in using optical properties to determine the existence of $\eta$-pairing in driven Hubbard chains. An alternate physical explanation that emerges from our study is that most of the reported superconductor-like optical properties do not in fact require the existence of superconducting order, but rather are a consequence of number-phase duality effects that are common between Mott insulators and superconductors. To establish the existence of superconductivity, additional evidence such as the existence of supercurrent (through Shapiro steps or otherwise) or other direct evidence of off-diagonal long-range order is required.

\begin{acknowledgments}
We thank S. Sankar for his valuable comments.
\end{acknowledgments}

\appendix
\begin{widetext}
\section{Derivation of the phase model}\label{app:model}

Here we sketch the steps leading from the microscopic Hamiltonian of Eq. (\ref{eq:modelH}) to the phase model. 
We first perform the Hubbard-Stratonovich decoupling of the Coulomb interaction: 
\begin{eqnarray}
e^{-i\int_{t}H_{C}} & = & \exp\left[-i\sum_{k}\int_{t}E_{C}\left(\sum_{\alpha}c^{\dagger}_{k,\alpha}c_{k,\alpha}-N_{0}\right)\left(\sum_{\alpha}c^{\dagger}_{k,\alpha}c_{k,\alpha}-N_{0}\right)\right]\nonumber \\
 & \propto & \int DV_k\exp\left[i\sum_{k}\int_{t}\frac{1}{4E_{C}}\left(V_k-2E_{C}\left(\sum_{\alpha}c^{\dagger}_{k,\alpha}c_{k,\alpha}-N_{0}\right)\right)^{2}\right]e^{-i\int_{t}H_{C}}
\end{eqnarray}

Since we are interested in the nonequilibrium response, we go over to the Keldysh action formalism, and label the fields on the forward and backward time contours respectively by superscripts $+$ and $-.$ The Hubbard-Stratonovich fields cause large shifting of the entire electron energy bands during tunneling events. To eliminate these fields we perform gauge transformations $c_{k}^{\pm}\rightarrow e^{-i\phi_{k}^{\pm}}c_{k}^{\pm}$ on the fermion fields such that $\partial_t \phi_{k}^{\pm}=V_{k}^{\pm}.$ While the Hubbard-Stranotonovich fields get eliminated from the single electron energies, they clearly appear now in the interdot tunneling terms.  It is convenient to perform here a basis rotation in Keldysh space so that the Green functions (see below) have the customary retarded, advanced, or Keldysh forms. To this end we introduce the ``classical'' ($c$) and ``quantum'' ($q$) components,
\begin{eqnarray}
V^{c} & = & \frac{1}{2}(V^{+}+V^{-})\mbox{ , }V^{q}=V^{+}-V^{-},\\
c^{c} & = & \frac{1}{\sqrt{2}}(c^{+}+c^{-})\mbox{ , }c^{q}=\frac{1}{\sqrt{2}}(c^{+}-c^{-}),\\
(c^{c})^{\dagger} & = & \frac{1}{\sqrt{2}}((c^{+})^{\dagger}-(c^{-})^{\dagger})\mbox{ , }(c^{q})^{\dagger}=\frac{1}{\sqrt{2}}((c^{+})^{\dagger}+(c^{-})^{\dagger}),\\
\Psi & = & \begin{pmatrix}c^{c}\\
c^{q}
\end{pmatrix}\mbox{ , }\Psi^{\dagger}=\begin{pmatrix}(c^{c})^{\dagger} & (c^{q})^{\dagger}\end{pmatrix}.
\end{eqnarray}

The Keldysh action $S=S_0 + S_C + S_{\text{tun}}$ now takes the form,
\begin{eqnarray}
S_{0} & = & \sum_{k,\alpha}\int_{t}\Psi_{k,\alpha}^{\dagger}\begin{bmatrix}i\partial_{t}+i\eta+\mu-\epsilon_{\alpha} & 2 i\eta F_{k}\\
0 & i\partial_{t}-i\eta+\mu-\epsilon_{\alpha}
\end{bmatrix}\Psi_{k,\alpha},\\
S_{C} & = & \sum_{k}\int_{t}\left(\frac{1}{2E_{c}}\partial_{t}\phi_{k}^{c}\partial_{t}\phi_{k}^{q}+N_{0}\partial_{t}\phi_{k}^{q}\right),\label{eq:SC-N0}\\
S_{\text{tun}} & = & \sum_{k,\alpha,\beta}\int_{t}\left(t^{\alpha,\beta}_{k,k+1}\bar{\Psi}_{k\alpha}\exp(-i\hat{\phi}_{k,1})\Psi_{k+1,\beta}+\mbox{c.c.}\right),
\end{eqnarray}
where $\hat{\phi}_{k,1}=\hat{\phi}_{k+1}-\hat{\phi}_{k},$ with $\hat{\phi}_k = \phi_{k}^{c}+(1/2)\phi_{k}^{q}\sigma_1.$
Here, $F_k (t,t')$ is related to the non interacting distribution function of electrons in the $k^{th}$ dot. If the number of dot electrons is large, then it can be shown \cite{S_Sankar_V_Tripathi} that $F_k$ can be approximated by its equilibrium value $F_k(t-t').$ In the frequency domain, this function is given by $F(\omega)=1-2f(\omega)$, where $f(\omega)$ is the Fermi-Dirac distribution function. The infinitesimally small positive constant $\eta$ has been added for the theory to have proper causal structure.

The fermion-bilinear part of the action, $S_F=S_0+S_{\text{tun}},$ can be integrated out easily. 
Let fermion Lagrangian density is $L_F=\Psi^\dagger \hat{G}^{-1} \Psi $, with
	\begin{equation}
		\hat{G}^{-1}=\hat{G}^{-1}_0 + \hat{T},
	\end{equation}
	where
	\begin{eqnarray}
		(\hat{G}_0)_{k,\alpha ; k,\alpha}^{-1} &=& \begin{pmatrix}
		(g^R_{k,\alpha})^{-1} & 2 \iota \eta F_k \\
		0 & (g^A_{k,\alpha})^{-1}
		\end{pmatrix},\\
		\hat{T}_{k,\alpha ; k+1,\beta} &=&  t^{\alpha\beta}_{k,k+1} e^{-i \hat{\Phi}_{k,1}}.
	\end{eqnarray}
	The diagonal elements are the inverse retarded and advanced Green functions,
	\begin{equation}
		(g^{R,A}_{k,\alpha})^{-1} = i \partial_t \pm i \eta +\mu-\epsilon_\alpha,
	\end{equation}
	respectively. The interdot hopping matrix $\hat{T}$ is diagonal in Keldysh space as well as in the time indices. We integrate out the fermions and get $ Z= \int D\phi \exp(i S_C[\phi]+ \text{tr}\ln(i \hat{G}^{-1} )),$, where we express the fermionic determinant as
	\begin{equation}
    \ln(\hat{G}^{-1}) = \ln(1 + \hat{G}_0 \hat{T}) + \ln(\hat{G}_0^{-1}).
	\end{equation}
	Now, to obtain the action in terms of the phase fields, we make a Taylor expansion of $\ln(1 + \hat{G}_0 \hat{T})$ in increasing powers of the tunneling. The first term clearly vanishes as $\hat{G}_0$ is diagonal in $k$ and $ T_{k;k}=0.$ 
	Up to second order in $\hat{T},$ we have
	\begin{equation}
		Z= \int D\phi \hspace{4pt} \exp\left( i S_C[\phi] + i S_{\text{tun}}[\phi] \right),\hspace{10pt} S_{\text{tun}}[\phi]= \frac{i}{2} \text{tr}(\hat{G}_0 \hat{T} \hat{G}_0 \hat{T}). 
	\end{equation}
	
	Here $\hat{G}_0$ has the following structure in the Keldysh space,
	\begin{equation}
		(\hat{G}_0)_{k,\alpha ; k,\alpha}(t,t') = \begin{pmatrix}
		g^R_{k,\alpha} &  F_k(g^R_{k,\alpha}-g^A_{k,\alpha} ) \\
		0 & g^A_{k,\alpha}
		\end{pmatrix} (t,t'),
	\end{equation}
	where
	\begin{equation}
		g^{R,A}_{k,\alpha}(t,t')= \frac{1}{2\pi} \int_\omega g^{R,A}_{k,\alpha}(\omega) e^{-i \omega (t-t')}=\int_\omega\frac{e^{-i \omega (t-t')}}{\omega \pm  i \eta +\mu-\epsilon_\alpha}.
	\end{equation}
	We assume the matrix elements of $\hat{T}$ are independent of the energy indices and further we replace the discrete summation over the energy indices by integrals, thereby, obtaining expression for $tr (\hat{G}_0 \hat{T} \hat{G}_0 \hat{T})$ in terms of $G^{R,A}(\omega)=\frac{1}{2\pi} \int_\epsilon g^{R,A}_{k,\epsilon}(\omega)$.
	Denoting the mean square tunneling matrix connecting pairs of levels in the neighbouring sites as $|t|^2,$ we obtain $S_{\text{tun}} \propto g=\pi^2 |t|^2/\delta^2 $ and given by,
	\begin{equation}
		S_{\text{tun}}[\phi]= g \sum_k \int_{t,t'} \begin{pmatrix}
		e^{-i\phi^+_{k,1}} & e^{-i\phi^-_{k,1}}
		\end{pmatrix}_t L_{k,1}(t-t') \begin{pmatrix}
		e^{i\phi^+_{k,1}}\\e^{i\phi^-_{k,1}}
		\end{pmatrix}_{t'}
	\end{equation}
	where,
	\begin{equation}
		L=\frac{1}{4} \begin{pmatrix}
		\Sigma^R+\Sigma^A+\Sigma^K & \Sigma^R-\Sigma^A-\Sigma^K\\ -\Sigma^R+\Sigma^A-\Sigma^K & -\Sigma^R-\Sigma^A+\Sigma^K
		\end{pmatrix}.
	\end{equation}
	The functions $\Sigma^{R,A}$ have a causal structure like the $G^{R,A}$ and are given by
	\begin{eqnarray}
		\Sigma^{R(A)}_{k,1}(t) &=& i \left( G^{R(A)}(t)G^{K}_k(-t)+G^{K}_{k+1}(t)G^{A(R)}(-t) \right),\\
		\Sigma^K_{k,1}(t)&=&i \left( G^{K}_k(-t)G^{K}_{k+1}(t)- (G^R-G^A)_t(G^R-G^A)_{-t} \right).
	\end{eqnarray}
	where the Keldysh component $G^K_k = F_k (G^R-G^A).$ Note that the kernel $L$ involves products of single-fermion Green functions, and thus represents a bosonic propagator. It takes a very simple form in the frequency domain. 

We make use of the following identities in the frequency domain to simplify the $L$ matrix elements:

\begin{eqnarray}
(G^R-G^A)_\epsilon=-i,\\
\int_\epsilon\frac{1}{2\pi}\left[F(\epsilon+\omega)-F(\epsilon)\right]=\dfrac{\omega}{\pi},\\
\int_\epsilon\frac{1}{2\pi}\left[1-F(\epsilon-\omega)F(\epsilon)\right]=\dfrac{\omega}{\pi}F_b(\omega),
\end{eqnarray}

where $F_b(\omega)=1+2f_b(\omega)$ and $f_b(\omega)$ is the equilibrium Bose-Einstein distribution function. Above identities enables us to make a simplification as follows:

\begin{eqnarray}
\left(\Sigma^R_{k,1}-\Sigma^A_{k,1}\right)_\omega=\dfrac{i}{\pi}\omega,\\
(\Sigma^K_{k,1})_\omega=\dfrac{i}{\pi}\omega F_b(\omega)
\end{eqnarray} 

We now justify dropping the higher order tunneling terms such as $\text{tr}(\hat{G}_0\hat{T}\hat{G}_0\hat{T}\hat{G}_0\hat{T}\hat{G}_0\hat{T})$ in the large-$\mathcal{N}$ approximation. Physically, the tunneling matrix elements $t_{kk'}^{\alpha\beta}$ must scale as $1/\mathcal{N}$ (where $\mathcal{N}$ is of the order of the number of conduction electrons in a dot) so that the dimensionless interdot conductance $g\sim |t|^2/\delta^2$ is independent of $\mathcal{N}$ and physically meaningful. Higher order terms do not contribute in the large $\mathcal{N}$ limit. For example, in the aforementioned fourth order term, the tunneling elements contribute an overall scaling factor of $1/\mathcal{N}^4$ while the sum over internal indices contributes only a scaling factor of $\mathcal{N}^3,$ resulting in this term becoming insignificant in the large-$\mathcal{N}$ sense. For a detailed analysis of the role of large-$\mathcal{N},$ we refer to the arguments given in Ref. \cite{S_Sankar_V_Tripathi}.
		
\end{widetext}

\bibliographystyle{apsrev4-1}

\end{document}